\documentclass{article}[9pt]
\usepackage{spconf,amsmath,epsfig}
\ninept

\title{Distributed conjugate gradient strategies for parameter
estimation over sensor networks
 \vspace{-0.35em}} \name{Songcen Xu and Rodrigo c. de Lamare
\vspace{-0.75em}}
\address{Communications Research Group, Department of
Electronics, University of York, U.K. \\
    Email: sx520@ohm.york.ac.uk, rcdl500@ohm.york.ac.uk  \vspace{-0.75em}
    }
\begin{document}
\maketitle
\begin{abstract}
This paper presents distributed adaptive algorithms based on the
conjugate gradient (CG) method for distributed networks. Both
incremental and diffusion adaptive solutions are all considered. The
distributed conventional (CG) and modified CG (MCG) algorithms have
an improved performance in terms of mean square error as compared
with least-mean square (LMS)-based algorithms and a performance that
is close to recursive least-squares (RLS) algorithms . The resulting
algorithms are distributed, cooperative and able to respond in real
time to changes in the environment. \vspace{-0.25em}
\end{abstract}

\vspace{-0.5em}

\section{Introduction}
In recent years, distributed processing has become popular in
wireless communication networks. This kind of processing can collect
data at each node in a given area and convey the information to the
whole network in a distributed way. For instance, each node can get
the information from its neighbors, and then combine it with the use
of distributed adaptive algorithms; each node has the ability to
estimate the nearby environment itself \cite{Lopes}. When compared
with the centralized solution, the distributed solution has
significant advantages. The centralized solution needs a central
processor, where each node sends its information to the central
processor and gets the information back after the processor
completes the task. This type of communication needs the central
processor to be powerful and reliable. With distributed solutions,
each node only requires the local information and its neighbors to
process the information. This kind of processing can significantly
reduce the amount of processing and the communications bandwidth
requirements.

There are three main cooperation modes: the incremental, the
diffusion, and the probabilistic diffusion modes \cite{Lopes}. For
the incremental mode, we can interpret it as a cycle in which the
information goes through the nodes in one direction, which means
each node passes the information to its adjacent node in a
pre-determined direction. Because of its simple method, the need for
communication and power is the least \cite{Lopes}. For the diffusion
mode, each node transfers information to its whole neighbors. This
kind of processing costs a huge amount of communication resources,
but each node will get more information. To avoid the high
communication cost, another kind of diffusion termed probabilistic
diffusion is used. Instead of transferring information to all its
neighbors, each node transfers data to a selected group of its
neighbors, which can be chosen randomly.

Several algorithms have already been developed and reported in the
literature for distributed networks. Steepest-descent, least mean
square (LMS) \cite{Lopes} and affine projection (AP) \cite{Li}
solutions have been considered with incremental adaptive strategies
over distributed networks \cite{Lopes}, while the LMS and recursive
least squares (RLS) algorithms have been reported using diffusion
adaptive strategies \cite{Lopes2,Cattivelli,Mateos}. Although the
LMS-based algorithms have their own advantages, when compared with
conjugate gradient (CG) algorithms, their shortages are obvious.
First, for the LMS- based algorithms, the adaptation speed is often
slow, especially for the conventional LMS algorithm. Second, when we
are trying to increase the adaptation speed, the system stability
may decrease significantly. Furthermore, the RLS-based algorithms
usually have a high complexity. In order to develop a set of
distributed solutions with a more attractive tradeoff between
performance and complexity, we focus on the CG algorithm. The CG
algorithm has a faster convergence rate \cite{Axelsson} than the
LMS-type algorithms and a lower computational complexity than
RLS-type techniques. In this paper, the main contribution is to
develop distributed CG algorithms for both incremental and diffusion
adaptive strategies. In particular, we develop distributed versions
of the conventional CG algorithm and of the modified CG algorithm
for use in distributed estimation over sensor networks.

This paper is organized as follows. Section 2 describes the system
model and states the problem. Section 3 presents the incremental
distributed CG algorithms, whereas Section 4 considers the diffusion
distributed CG algorithms. Section 5 presents and discusses the
simulation results, whereas Section 6 gives the conclusions.

\section{System model and problem statement}
In this part, we describe a system model of the distributed
estimation scheme over sensor networks and introduce the
problem statement.

\subsection{System model}
The basic idea of this system model is that for each node in a sensor network a designer deals with
 a system identification problem. Each node is equipped
with an adaptive filter. We focus on the processing of an adaptive
filter for adjusting the weight vector ${\boldsymbol \omega}_o$ with
coefficients $\boldsymbol \omega_k$ ($k=1,\ldots, M$), where M is the length of
the filter. The desired signal of each node at time instant $ i $ is
\begin{equation}
{d^{(i)}} = {\boldsymbol {\omega}}_0^H{\boldsymbol x^{(i)}} +{n^{(i)}},~~~
i=1,2, \ldots, N \label{obsig},
\end{equation}
where $d^{(i)}$ is the received signal sample, ${\boldsymbol x^{(i)}}$
is the $M \times 1$ input signal vector, ${\boldsymbol {\omega}}_0$
is the $M \times 1$ system weight vector, ${ n^{(i)}}$ is the noise
sample at each receiver, $(\cdot)^H$ denotes Hermitian transpose and
$N$ is the number of time instants. At the same time, the output of
the adaptive filter for each node is given by
\begin{equation}
{y^{(i)}} = {\boldsymbol {\omega}^{(i)}}^H{\boldsymbol x^{(i)}},~~~ i=1,2,
\ldots, N \label{obsig},
\end{equation}
where ${\boldsymbol \omega^{(i)}}$ is the local estimator ${\boldsymbol \omega}$
for each node at time instant $ i $.

\subsection{Problem statement}
To get the optimum solution of the weight vector, we need
to solve a problem expressed in the form of a minimization of a cost function.
Consider a network which has $ N $ nodes, each node $ k $ has access
to time realizations \{${{d_k^{(i)},{\boldsymbol u}_k^{(i)}}}$\} of zero-
mean spatial data \{${{ d_k,\boldsymbol u_k}}$\},~~~ $k$=1,2,
\ldots, N, where each ${{ d_k}}$ is a scalar measurement and each
${{\boldsymbol u_k}}$ is a row regression vector \cite{Lopes}. After
that, two global matrices are built which are used to collect the
measurement data and regression data that are expressed in the form
of the matrices:
\begin{equation}
{\boldsymbol X} = [{\boldsymbol x_1,\boldsymbol x_2,...\boldsymbol
x_N} ],~  ({N \times M}) \label{obsig}
\end{equation}
\begin{equation}
{\boldsymbol d} = [ d_1, d_2, \ldots  d_N]^T,~ ({N \times
1})\label{obsig}
\end{equation}
The data which these two equations collect cover all nodes. In order
to design an algorithm to compute the optimum estimation value, we
need to first define a cost function:
\begin{equation}
{J({\boldsymbol \omega})} = {\boldsymbol E [||{\boldsymbol d}-
{\boldsymbol X}{\boldsymbol \omega}||^2}] \label{obsig2},
\end{equation}
where the ${J({\boldsymbol \omega})}$ is used to calculate the MSE
and our aim is to minimize the cost function. The optimal solution
should satisfy \cite{Lopes}:
\begin{equation}
{E}[{\boldsymbol X^H}({\boldsymbol d - {\boldsymbol X}{\boldsymbol
\omega}_o})] = {\boldsymbol 0}. \label{obsig}
\end{equation}
Meanwhile, the ${\boldsymbol \omega}_o $ is also the solution to:
\begin{equation}
{\boldsymbol b} = {\boldsymbol R} {\boldsymbol \omega}_{o}
\label{obsig3},
\end{equation}
where the $M\times M$ autocorrelation matrix is given by
${\boldsymbol R}=E[{\boldsymbol X}^H {\boldsymbol X}]$ and
${\boldsymbol b}=E[{\boldsymbol X}^H {\boldsymbol d}]$ is an $M
\times 1$ cross-correlation matrix. In this work, we focus on
incremental and diffusion CG-based algorithms to solve the equation
and perform estimation in a distributed fashion.

\section{Proposed Incremental Distributed CG - Based Algorithms}

For distributed estimation over sensor networks, we develop two CG-
based algorithms which are the CCG and MCG with incremental
distributed solution (IDCG). We the derive the CG- based algorithms
first, then we devise distributed versions of these algorithm for
use in the network in an incremental and distributed way.

\subsection{Derivation of the CG- based algorithms}

When a CG algorithm is used in adaptive signal processing, it solves
the following equation\cite{Chang}:
\begin{equation}
{\boldsymbol R_k^{(i)}(j)} {\boldsymbol \omega_k^{(i)}(j)}= {\boldsymbol b_k^{(i)}(j)} \label{obsig},
\end{equation}
where  ${\boldsymbol R_k^{(i)}(j)}$ is the $M \times M$ correlation
matrix for the input data vector, and ${\boldsymbol b_k^{(i)}(j)}$
is the $M \times 1$ cross-correlation vector between the input data
vector and $d$ is the desired response. To solve this equation, we
need to obtain:
\begin{equation}
{\boldsymbol \omega_k^{(i)}(j)} = [{\boldsymbol R_k^{(i)}(j)}]^{-1} {\boldsymbol b_k^{(i)}(j)} \label{obsig}.
\end{equation}
In the CG- based algorithm, the iteration procedure is introduced. For the $j$th iteration, we choose the
negative direction as:
\begin{equation}
{\boldsymbol g_k^{(i)}(j)}={\boldsymbol b_k^{(i)}(j)} - {\boldsymbol R_k^{(i)}(j)}{\boldsymbol \omega} \label{obsig}.
\end{equation}
The CG-based weight vector $\boldsymbol \omega_k^{(i)}(j)$ is defined as:
\begin{equation}
{\boldsymbol \omega_k^{(i)}(j)} = {\boldsymbol \omega_k^{(i)}(j-1)} + {\alpha_k^{(i)}(j)}{\boldsymbol p_k^{(i)}(j)} \label{obsig},
\end{equation}
where $\boldsymbol p(j)$ is the direction vector with conjugacy and
$\alpha(j)$ is calculated by replacing (11) in (9), then taking the
gradient with respect to $\alpha(j)$ and using (11), we get:
\begin{equation}
{\alpha_k^{(i)}(j)} = \frac{\rho_k^{(i)}(j-1)}
{{\boldsymbol p_k^{(i)}(j)}^H{\boldsymbol c_k^{(i)}(j)}} \label{obsig},
\end{equation}
where
\begin{equation}
{\rho_k^{(i)}(j)} = {\boldsymbol g_k^{(i)}(j)}^H {\boldsymbol
g_k^{(i)}(j)}
\end{equation}
and
\begin{equation}
{\boldsymbol c_k^{(i)}(j)} = {\boldsymbol R_k^{(i)}(j)} {\boldsymbol
p_k^{(i)}(j)} .
\end{equation}
The direction vector $\boldsymbol p_k^{(i)}(j)$ in (11) is defined as:
\begin{equation}
{\boldsymbol p_k^{(i)}(j+1)} = {\boldsymbol g_k^{(i)}(j)} + {\beta_k^{(i)}(j)}{\boldsymbol p_k^{(i)}(j)} \label{obsig},
\end{equation}
where $\beta_k^{(i)}(j)$ is calculated by the Gram Schmidt
orthogonalization procedure\cite{Golub} for the conjugacy:
\begin{equation}
{\beta_k^{(i)}(j)} = \frac{ \rho_k^{(i)}(j)}
{\rho_k^{(i)}(j-1)} \label{obsig}
\end{equation}
with
\begin{equation}
{\rho_k^{(i)}(j)} = {\boldsymbol g_k^{(i)}(j)}^H {\boldsymbol
g_k^{(i)}(j)} .
\end{equation}
Besides the basic CG algorithm, there are two ways to define the
correlation and cross-correlation matrices which are 'finite sliding
data window' and 'exponentially decaying data window'\cite{Chang}.
In this paper, we mainly focus on the 'exponentially decaying data
window' because this approach employs the same correlation matrix as
the RLS algorithm. The recursions are given by:
\begin{equation}
{\boldsymbol R_k^{(i)}}(j) = {\lambda_f}{\boldsymbol R_{k-1}^{(i)}}(j) + {\boldsymbol
x_k^{(i)}}(j)[{\boldsymbol x_k^{(i)}}(j)]^H \label{obsig}
\end{equation}
\begin{equation}
{\boldsymbol b_k^{(i)}}(j) = {\lambda_f}{\boldsymbol b_{k-1}^{(i)}}(j) +
d_k^{(i)*}(j){\boldsymbol x_k^{(i)}}(j) \label{obsig}
\end{equation}
where $\lambda_f$ is the forgetting factor.

\subsection{Incremental Distributed CG - Based Solutions}

\begin{figure}[!htb]
\begin{center}
\def\epsfsize#1#2{0.825\columnwidth}
\epsfbox{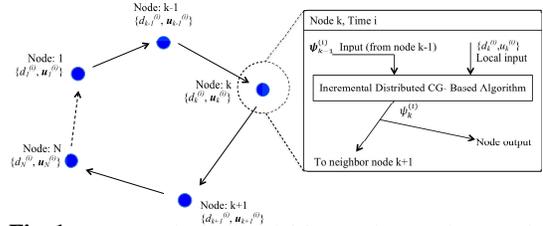} \vspace{-1.85em} \caption{\footnotesize
Incremental Distributed CG- Based Network Processing}\vspace{-0.5em}
\label{fig1}
\end{center}
\end{figure}

In the incremental distributed model of our algorithm, each node is
only allowed to communicate with its direct neighbor at each time
instant. To describe the whole process,we define a cycle, where each
node in this network could only access its immediate neighbor in
this cycle \cite{Lopes}. The quantity $\boldsymbol\psi_k ^{(i)}$ is
defined as a local estimate of $\boldsymbol\omega^o$ at time $i$. As
a result, we assume that node $k$ has access to an estimate of
$\boldsymbol\omega^o$  at its immediate neighbor node $k-1$ which is
$\boldsymbol\psi_{k-1} ^{(i)}$ in the defined cycle. Fig.\ref{fig1}
shows its processing.

Based on the main steps of the CG algorithm, we propose two
distributed adaptive filtering algorithms, namely, the CCG and the
MCG for distributed estimating over sensor networks. The difference
between these two strategies is that the CCG needs to run $k$
iterations while the MCG only needs one iteration. The
implementation of Incremental Distributed CCG Solution (IDCCG) is
showed in Table \ref{table1}. Similarly to CCG algorithm
lowercase,the Incremental Distributed MCG solution (IDMCG) only
needs one iteration per time instant and the details are shown in
Table \ref{table1}.
\begin{table*}
\caption{IDCG Solutions} \centering
\begin{tabular}{l|l}\hline
IDCCG Solution&IDMCG Solution\\
\hline
Initialization:&Initialization:
\\
$\boldsymbol\omega_0=0, \boldsymbol g(0)=\boldsymbol b, \boldsymbol p(1)=\boldsymbol g(0)$&$\boldsymbol\omega_0=0, \boldsymbol g_0=\boldsymbol b, \boldsymbol p_1=\boldsymbol g_0$
\\
For each time instant $i$=1,2, \ldots, n&For each time instant $i$=1,2, \ldots, n
\\
For each node $k$=1,2, \ldots, N&For each node $k$=1,2, \ldots, N
\\
$\boldsymbol\psi_0 ^{(i)}(1) = \boldsymbol\omega_{i-1} \notag$&${\boldsymbol R_k^{(i)}} = {\lambda_f}{\boldsymbol R_{k-1}^{(i)}} + {\boldsymbol x_k^{(i)}}{ {\boldsymbol x_k^{(i)}}^H} \notag$
\\
For iterations $j$=1,2, \ldots, J&${\boldsymbol b_k^{(i)}} = {\lambda_f}{\boldsymbol b_{k-1}^{(i)}} + { d_k^{(i)}}{\boldsymbol x_k^{(i)}} \notag$
\\
${\boldsymbol R_k^{(i)}}(j) = {\lambda_f}{\boldsymbol R_{k-1}^{(i)}}(j) + {\boldsymbol x_k^{(i)}}(j){ {\boldsymbol x_k^{(i)}}^H}(j) \notag$&$\boldsymbol \psi_0 ^{(i)} = \boldsymbol \omega_{i-1} \notag$
\\
${\boldsymbol b_k^{(i)}}(j) = {\lambda_f}{\boldsymbol b_{k-1}^{(i)}}(j) + { d_k^{(i)}}(j){\boldsymbol x_k^{(i)}}(j) \notag$&${\alpha_k^{(i)}} = {\eta}[{{\boldsymbol p_k^{(i)}}^H}{\boldsymbol R_k^{(i)}}{\boldsymbol p_k^{(i)}}]^{-1}[{{\boldsymbol p_{k}^{(i)}}^H}{\boldsymbol g_{k-1}^{(i)}}]\notag$
\\
${\alpha_k^{(i)}(j)} = {\eta}[{\boldsymbol p_k^{(i)}(j)^H}{\boldsymbol R_k^{(i)}(j)}{\boldsymbol p_k^{(i)}(j)}]^{-1}[{\boldsymbol p_{k}^{(i)}(j)^H}{\boldsymbol g_{k-1}^{(i)}(j)}]\notag$& where ($\lambda_f -0.5)\leq\eta\leq\lambda_f$
\\
where ($\lambda_f -0.5)\leq\eta\leq\lambda_f $&${\boldsymbol \psi_k^{(i)}} = {\boldsymbol \psi_{k-1}^{(i)}} + { \alpha_k^{(i)}}{\boldsymbol p_k^{(i)}} \notag$
\\
${\boldsymbol \psi_k^{(i)}(j)} = {\boldsymbol \psi_{k-1}^{(i)}(j)} + {\alpha_k^{(i)}(j)}{\boldsymbol p_k^{(i)}(j)} \notag$&${\boldsymbol g_k^{(i)}}= {\lambda_f}{\boldsymbol g_{k-1}^{(i)}} - {\alpha_k^{(i)}}{\boldsymbol R_k^{(i)}}{\boldsymbol p_k^{(i)}}\notag$
\\
${\boldsymbol g_k^{(i)}(j)}= {\boldsymbol g_k^{(i)}(j-1)} - {\alpha_k^{(i)}(j)}{\boldsymbol R_k^{(i)}(j)}{\boldsymbol p_k^{(i)}(j)}\notag$&${\beta_k^{(i)}} = [{{\boldsymbol g_{k-1}^{(i)}}^H}{\boldsymbol g_{k-1}^{(i)}}]^{-1}[({\boldsymbol g_k^{(i)}-\boldsymbol g_{k-1}^{(i)}})^H\boldsymbol g_k^{(i)}]
\notag$
\\
${\beta_k^{(i)}(j)} = [{{\boldsymbol g_k^{(i)}(j-1)}^H}{\boldsymbol g_k^{(i)}(j-1)}]^{-1}[{{(\boldsymbol g_k^{(i)}(j)})^H}\boldsymbol g_k^{(i)}(j)]\notag$&${\boldsymbol p_{k+1}^{(i)}} = {\boldsymbol g_k^{(i)}} + { \beta_k^{(i)}}{\boldsymbol p_k^{(i)}}+ {\boldsymbol x_k^{(i)}}[{ d_k^{(i)}}-{ {\boldsymbol x_k^{(i)}}^H}{\boldsymbol \psi_{k-1}^{(i)}}] \notag$
\\
${\boldsymbol p_k^{(i)}(j+1)} = {\boldsymbol g_k^{(i)}(j)} + { \beta_k^{(i)}(j)}{\boldsymbol p_k^{(i)}(j)} \notag$&$ \boldsymbol\omega_i =\boldsymbol \psi_N ^{(i)} \notag$
\\
$j=j+1 \notag$&$k=k+1 $
\\
After J iterations&After N iterations
\\
$k=k+1 \notag$&$i=i+1 \notag$
\\
After N iterations&
\\
$\boldsymbol\omega_i = \boldsymbol\psi_N ^{(i)} \notag$&
\\
$i=i+1 \notag$&
\\\hline
\end{tabular}
\label{table1}
\end{table*}
These two Incremental Distributed CG- Based Solutions can be
summarised as:
\begin{description}
 \item[1)]
 assess local error
 \item[2)]
 update its weight vector
 \item[3)]
 pass the updated weight estimate  $\boldsymbol \psi_k ^{(i)}$ to its neighbor node
\end{description}
The idea of the MCG algorithm comes from the CCG algorithm. Instead of equation (10), we redefine the negative gradient vector with a recursive expression\cite{Wang}:
\begin{equation}
\begin{split}
{\boldsymbol g_k^{(i)}} &= {\boldsymbol b_k^{(i)}} - {\boldsymbol R_k^{(i)}}{\boldsymbol \omega_k^{(i)}} \\
& = {\lambda_f}{\boldsymbol g_{k-1}^{(i)}} - {\alpha_k^{(i)}}{\boldsymbol R_k^{(i)}}{\boldsymbol p_k^{(i)}}
\\
& \ \ \ \ + {\boldsymbol x_k^{(i)}}[{d_k^{(i)}}-{\boldsymbol x_k^{(i)}}^H{\boldsymbol \omega_{k-1}^{(i)}}]\label{obsig}.
\end{split}
\end{equation}
Premultiplying (17) by $\boldsymbol p_k^H$ and considering
$\boldsymbol p_k$ uncorrelated with $\boldsymbol x_k$, $ d_k$ and
$\boldsymbol \omega_{k-1}$ and then taking the expectation, we get:
\begin{equation}
\begin{split}
E[{\boldsymbol p_k^{(i)}}^H{\boldsymbol g_k^{(i)}}] & = {\lambda_f}E[{\boldsymbol p_k^{(i)}}^H{\boldsymbol g_{k-1}^{(i)}}]
\\
& \ \ \ \ - E[{\alpha_k^{(i)}}]E[{\boldsymbol p_k^{(i)}}^H{\boldsymbol R_k^{(i)}}{\boldsymbol p_k}^{(i)}]
\\
& \ \ \ \ + E[{\boldsymbol p_k^{(i)}}^H{\boldsymbol x_k^{(i)}}[{d_k^{(i)}}-{\boldsymbol x_k^{(i)}}^H{\boldsymbol \omega_{k-1}^{(i)}}]]\label{obsig}.
\end{split}
\end{equation}
Assuming the algorithm converges, then the last term of (18) could be neglected and we will get:
\begin{equation}
E[{\alpha_k}^{(i)}] = \frac{E[{\boldsymbol p_{k}^{(i)}}^H{\boldsymbol g_{k}}^{(i)}]-{\lambda_f}E[{\boldsymbol p_{k}^{(i)}}^H{\boldsymbol g_{k-1}^{(i)}}]}
{E[{\boldsymbol p_k^{(i)}}^H{\boldsymbol R_k^{(i)}}{\boldsymbol p_k^{(i)}}]} \notag
\end{equation}
and
\begin{equation}
\begin{split}
({\lambda_f-0.5})\frac{E[{\boldsymbol p_{k}^{(i)}}^H{\boldsymbol g_{k-1}^{(i)}}]}
{E[{\boldsymbol p_k^{(i)}}^H{\boldsymbol R_k^{(i)}}{\boldsymbol p_k^{(i)}}]}
\leq E[{\alpha_k^{(i)}}]
\leq \frac{E[{\boldsymbol p_{k}^{(i)}}^H{\boldsymbol g_{k-1}^{(i)}}]}
{E[{\boldsymbol p_k^{(i)}}^H{\boldsymbol R_k^{(i)}}{\boldsymbol p_k^{(i)}}]}
\end{split}
\end{equation}
The inequalities in (19) are satisfied if we define:
\begin{equation}
{\alpha_k^{(i)}} = {\eta}\frac{{\boldsymbol p_{k}^{(i)}}^H{\boldsymbol g_{k-1}^{(i)}}}
{{\boldsymbol p_k^{(i)}}^H{\boldsymbol R_k^{(i)}}{\boldsymbol p_k^{(i)}}},
\end{equation}
where ($\lambda_f -0.5)\leq\eta\leq\lambda_f $.
The direction vector $\boldsymbol p_k$ is defined by:
\begin{equation}
{\boldsymbol p_{k+1}^{(i)}} = {\boldsymbol g_k^{(i)}} + {\beta_k^{(i)}}{\boldsymbol p_k^{(i)}}
\end{equation}
where $\beta_k$ is computed to avoid the residue produced by using the Polak- Ribiere approach \cite{Chang} which is given by:
\begin{equation}
{\beta_k^{(i)}} = \frac{({\boldsymbol g_k^{(i)}-\boldsymbol g_{k-1}^{(i)})^H\boldsymbol g_k^{(i)}}}
{{{\boldsymbol g_{k-1}^{(i)}}^H}{\boldsymbol g_{k-1}^{(i)}}}.
\end{equation}

\subsection{Computational Complexity}

To analyse the proposed incremental distributed CG algorithms, we
detail the computational complexity. Additions and multiplications
are used to measure the complexity and listed in Table \ref{table2}.
It is obvious that the complexity of the incremental distributed CCG
algorithm depends on the iteration number $j$.

\begin{table}
\centering \caption{Computational Complexity of Algorithms.}
\begin{tabular}{|c|c|c|}
\hline
Algorithm&Additions&Multiplications\\
\hline
IDCCG&$m^2+2m-2$&$m^2+3m$\\
&$+J(2m^2+7m-2)$&$J(3m^2+6m-2)$\\
\hline
IDMCG&$3m^2+11m-5$&$4m^2+11m-2$\\
\hline
IDLMS&$4m-1$&$3m+1$\\
\hline
IDRLS&$m^2+4m-1$&$m^2+5m$\\
\hline
\end{tabular}
\label{table2}
\end{table}

\section{Proposed Diffusion Distributed CG - Based Algorithms}

\subsection{Network Structure}
\begin{figure}[!htb]
\begin{center}
\def\epsfsize#1#2{0.85\columnwidth}
\epsfbox{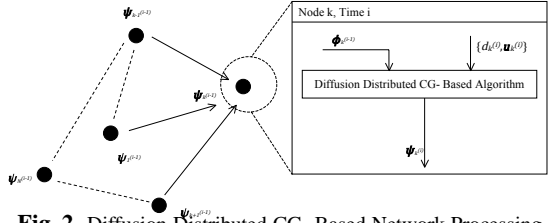} \vspace{-1.85em} \caption{\footnotesize Diffusion
Distributed CG- Based Network Processing}\vspace{-0.5em}
\label{fig2}
\end{center}
\end{figure}

For the diffusion distributed CG- based strategy, we consider a network structure where each node from the same neighborhood could exchange information with each other at every iteration. For each node in the network, it can collect information from all its neighbors and itself, and then convey all the information to its local adaptive filter and update the estimation of the weight vector through our algorithms. Specifically, at any time instant $i-1$, we define that node $k$ has access to a set of unbiased estimates $\{\boldsymbol\psi_k^{(i-1)}\}_{k\in N_k}$ from its neighborhood $N_k$ including itself. Then, these local estimates are combined at node $k$ as
\begin{equation}
{\boldsymbol\phi_k^{(i-1)}}  =  \sum_{l\in N_{k,i-1}} c_{kl} \boldsymbol\psi_l^{(i-1)}\label{obsig}
\end{equation}
where $c_{kl}$ should be satisfied
\begin{equation}
\sum_{l} c_{kl} =1 , l\in N_{k,i-1} \forall k\label{obsig}
\end{equation}
Among the strategies to choose the combiner C are the Metropolis, the Laplacian and the nearest neighbor rules\cite{Lopes1}. For our proposed diffusion distributed CG- based algorithm, we choose the Metropolis whose processing is shown in Fig.\ref{fig2} and operates as follows:
\begin{equation}
\left\{\begin{array}{ll}
c_{kl}= \frac{1}
{max(n_k,n_l)},\ \ \ \ \ \ \ \ \  \ \ \ \ \ \ \ \ \ \ \ \ \ \ \ \ \ \ \     $if\  $k\neq l$\  \ are\  linked$\\
c_{kl}=0,              \ \ \ \ \ \ \ \ \ \  \ \ \ \ \ \ \ \ \ \ \ \ \ \ \ \ \ \ \ \ \ \ \ \ \ \ \ \ \ \ \ \ \ \     $for\  $k$\  and\  $l$\ not\  linked$\label{obsig}\\
c_{kk} = 1 - \sum_{l\in N_k / k} c_{kl}, \ \ \ \ \ \ \ \ \ \ \ \ \ \ \ \ \ $for\  $k$\ =\ $l$$
\end{array}
\right.
\end{equation}

\subsection{Diffusion Distributed CG - Based Solutions}

The CCG and MCG algorithms are also developed for the diffusion
distributed CG - based solutions, the details for these two
algorithms are shown in Table \ref{table3}. To derive these two
algorithms, we first use equation (23) to get the unbiased estimates
$\boldsymbol\phi_k^{(i-1)}$ and substitute them into equation (11),
which results in:
\begin{equation}
{\boldsymbol \psi_k^{(i)}(j)} = {\boldsymbol \phi_{k}^{(i-1)}(j)} + {\alpha_k^{(i)}(j)}{\boldsymbol p_k^{(i)}(j)}
\end{equation}
The rest of derivation is similar to the incremental CG- based
solutions.

\begin{table*}
\centering \caption{DDCG Solutions}
\begin{tabular}{l|l}
\hline
DDCCG Solution&DDMCG Solution\\
\hline
Initialization:&Initialization:
\\
$\boldsymbol\omega_0=0, \boldsymbol g(0)=\boldsymbol b, \boldsymbol p(1)=\boldsymbol g(0)$&$\boldsymbol\omega_0=0, \boldsymbol g_0=\boldsymbol b,\boldsymbol p_1=\boldsymbol g_0$
\\
For each time instant $i$=1,2, \ldots, n&For each time instant $i$=1,2, \ldots, n
\\
For each node $k$=1,2, \ldots, N&For each node $k$=1,2, \ldots, N
\\
$\boldsymbol \phi_k ^{(-1)}(1) = 0 \notag$&${\boldsymbol R_k^{(i)}} = {\lambda_f}{\boldsymbol R_{k-1}^{(i)}} + {\boldsymbol x_k^{(i)}}{ {\boldsymbol x_k^{(i)}}^H} \notag$
\\
For iterations $j$=1,2, \ldots, J&${\boldsymbol b_k^{(i)}} = {\lambda_f}{\boldsymbol b_{k-1}^{(i)}} + { d_k^{(i)}}{\boldsymbol x_k^{(i)}} \notag$
\\
${\boldsymbol R_k^{(i)}}(j) = {\lambda_f}{\boldsymbol R_{k-1}^{(i)}}(j) + {\boldsymbol x_k^{(i)}}(j){ {\boldsymbol x_k^{(i)}}^H}(j) \notag$&$\boldsymbol\phi_k ^{(-1)}(1) = 0 \notag$
\\
${\boldsymbol b_k^{(i)}}(j) = {\lambda_f}{\boldsymbol b_{k-1}^{(i)}}(j) + {d_k^{(i)}}(j){\boldsymbol x_k^{(i)}}(j) \notag$&${\alpha_k^{(i)}} = {\eta}[{{\boldsymbol p_k^{(i)}}^H}{\boldsymbol R_k^{(i)}}{\boldsymbol p_k^{(i)}}]^{-1}[{{\boldsymbol p_{k}^{(i)}}^H}{\boldsymbol g_{k-1}^{(i)}}]\notag$
\\
${\alpha_k^{(i)}(j)} = {\eta}[{\boldsymbol p_k^{(i)}(j)^H}{\boldsymbol R_k^{(i)}(j)}{\boldsymbol p_k^{(i)}(j)}]^{-1} [{\boldsymbol p_{k}^{(i)}(j)^H}{\boldsymbol g_{k}^{(i)}(j-1)}]\notag$&where ($\lambda_f -0.5)\leq\eta\leq\lambda_f $
\\
where ($\lambda_f -0.5)\leq\eta\leq\lambda_f $&${\boldsymbol\phi_k^{(i-1)}}  =  \sum_{l\in N_{k,i-1}} c_{kl} \boldsymbol\psi_l^{(i-1)}\notag$
\\
${\boldsymbol\phi_k^{(i-1)}(j)}  =  \sum_{l\in N_{k,i-1}} c_{kl} \boldsymbol\psi_l^{(i-1)}(j)\notag$&${\boldsymbol \psi_k^{(i)}} = {\boldsymbol \phi_{k}^{(i-1)}} + {\alpha_k^{(i)}}{\boldsymbol p_k^{(i)}} \notag$
\\
${\boldsymbol \psi_k^{(i)}(j)} = {\boldsymbol \phi_{k}^{(i-1)}(j)} + {\alpha_k^{(i)}(j)}{\boldsymbol p_k^{(i)}(j)} \notag$&${\boldsymbol g_k^{(i)}}= {\lambda_f}{\boldsymbol g_{k-1}^{(i)}} - {\alpha_k^{(i)}}{\boldsymbol R_k^{(i)}}{\boldsymbol p_k^{(i)}} + {\boldsymbol x_k^{(i)}}[{d_k^{(i)}}-{ {\boldsymbol x_k^{(i)}}^H}{\boldsymbol \phi_{k}^{(i-1)}}]\notag$
\\
${\boldsymbol g_{k}^{(i)}(j)}= {\boldsymbol g_{k}^{(i)}(j-1)} - {\alpha_k^{(i)}(j)}{\boldsymbol R_k^{(i)}(j)}{\boldsymbol p_k^{(i)}(j)}\notag$&${\beta_k^{(i)}} = [{{\boldsymbol g_{k-1}^{(i)}}^H}{\boldsymbol g_{k-1}^{(i)}}]^{-1} [({\boldsymbol g_k^{(i)}-\boldsymbol g_{k-1}^{(i)}})^H\boldsymbol g_k^{(i)}]
\notag$
\\
${\beta_k^{(i)}(j)} = [{{\boldsymbol g_{k}^{(i)}(j-1)}^H}{\boldsymbol g_{k}^{(i)}(j-1)}]^{-1}[{{(\boldsymbol g_{k}^{(i)}(j)})^H}\boldsymbol g_{k}^{(i)}(j)]\notag$&${\boldsymbol p_{k+1}^{(i)}} = {\boldsymbol g_k^{(i)}} + { \beta_k^{(i)}}{\boldsymbol p_k^{(i)}} \notag$
\\
${\boldsymbol p_{k}^{(i)}(j+1)} = {\boldsymbol g_{k}^{(i)}(j)} + {\beta_k^{(i)}(j)}{\boldsymbol p_k^{(i)}(j)} \notag$&$k=k+1 \notag$
\\
$j=j+1 \notag$&After N iterations
\\
After J iterations&$\boldsymbol\omega_i =\boldsymbol\psi_N ^{(i)} \notag$
\\
$k=k+1 \notag$&$i=i+1 \notag$
\\
After N iterations
\\
$\boldsymbol \omega_i =\boldsymbol \psi_N ^{(i)} \notag$
\\
$i=i+1 \notag$
\\\hline
\end{tabular}
\label{table3}
\end{table*}

\subsection{Computational Complexity}
The computational complexity is used to analyse the proposed
diffusion distributed CG - based algorithms where additions and
multiplications are measured. The details are listed in Table
\ref{table4}. Similarly to incremental distributed CG - based
algorithms, it is clear that the complexity of the incremental
distributed CCG algorithm depends on the iteration number $j$ and
the number of linked nodes $l$.
\begin{table}
\centering
\caption{Computational Complexity Algorithms}
\begin{tabular}{|c|c|c|}
\hline
Algorithm&Additions&Multiplications\\
\hline
DDCCG&$m^2+2m-2$&$m^2+3m$\\
&$+J(2m^2+7m-2)$&$+J(3m^2+6m-2)$\\
&$+Lm$&$+Lm$\\
\hline
DDMCG&$3m^2+11m-5$&$4m^2+11m-2$\\
&$+Lm$&$+Lm$\\
\hline
DDLMS&$4m-1+Lm$&$3m+1+Lm$\\
\hline
DDRLS&$m^2+4m-1+Lm$&$m^2+5m+Lm$\\
\hline
\end{tabular}
\label{table4}
\end{table}

\section{Simulation Results}

In this part, we test the proposed incremental and diffusion
distributed CG - based algorithms in a sensor network and compare
the results with LMS, RLS and AP \cite{Li} algorithms based on the
performance of excess MSE (EMSE). For each test, the number of
repetitions is set to 1000, and we assume there are 20 nodes in the
network. The number of taps of the adaptive filter is 10, the
variance for the input signal and the noise are 1 and 0.001,
respectively. Besides, the noise samples are modelled as complex
Gaussian noise.

\subsection{Performance of Proposed IDCG Algorithms}

\begin{figure}[!htb]
\begin{center}
\def\epsfsize#1#2{0.825\columnwidth}
\epsfbox{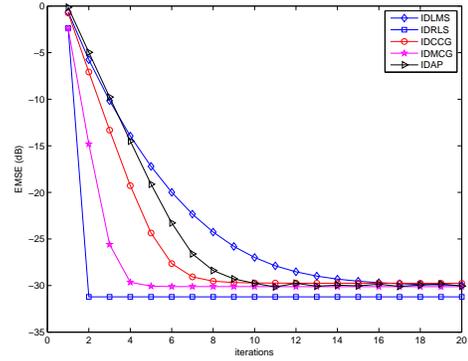} \vspace{-1.85em} \caption{\footnotesize Output
EMSE against the number of iterations for Incremental Strategy with
$\alpha_{IDLMS}$=0.005, $\lambda$=0.2,
$\lambda_{f-IDCCG}$=$\lambda_{f-IDMCG}$=0.25,
$\eta_{f-IDCCG}$=$\eta_{f-IDMCG}$=0.15, $j$=5, $\alpha_{IDAP}$=0.06,
$K$=2}\vspace{-0.5em} \label{fig3}
\end{center}
\end{figure}

First, we give out the definitions of the parameters of our test for
each algorithms and the network. After 1000 iterations, the
performance of each algorithm has been showed in Fig. \ref{fig3}. We
can see that, the performance of the IDMCG and IDCCG algorithm is
better than IDLMS, while IDMCG is very close to the IDRLS
algorithm's curve. The reason why the proposed IDMCG algorithm has a
better performance is IDMCG defined the negative gradient vector
$\boldsymbol g_{k}$ with a recursive expression and the $\beta_{k}$
is computed to avoiding the residue produced by using the Polak-
Ribiere approach. Comparing with the IDCCG algorithm, the IDMCG is a
non-reset and low complexity algorithm with one iteration per time
instant. Because of how often the algorithm is rest will influence
the performance, the IDMCG introduce the non-rest method together
with Polak- Ribiere method which is used for improved its
performance \cite{Chang}.

\subsection{Performance of Proposed DDCG Algorithms}

\begin{figure}[!htb]
\begin{center}
\def\epsfsize#1#2{0.825\columnwidth}
\epsfbox{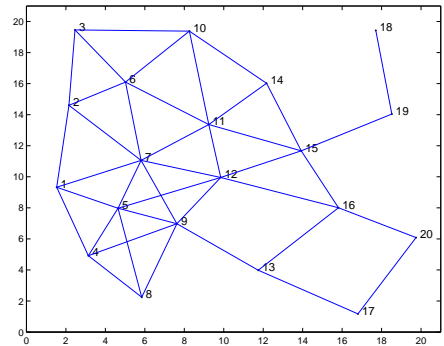} \vspace{-1.85em} \caption{\footnotesize Network
Structure}\vspace{-0.5em} \label{fig4}
\end{center}
\end{figure}

\begin{figure}[!htb]
\begin{center}
\def\epsfsize#1#2{0.825\columnwidth}
\epsfbox{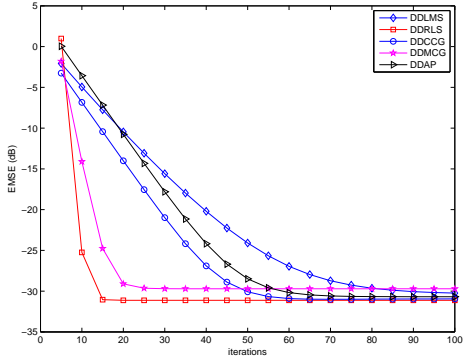} \vspace{-1.85em} \caption{\footnotesize Output
EMSE against the number of iterations for Diffusion Strategy with
$\alpha_{DDLMS}$=0.0075, $\lambda$=0.998, $\lambda_{f-DDCCG}$=0.25,
$\eta_{f-DDCCG}$=0.25, $j$=5, $\lambda_{f-DDMCG}$=0.46,
$\eta_{f-DDMCG}$=0.45, $\alpha_{DDAP}$=0.075, $K$=2}\vspace{-0.5em}
\label{fig5}
\end{center}
\end{figure}

For this group of Proposed DDCG Algorithms' test, we use some
similar definitions of parameters as in the last part. For the
diffusion strategy, we build the link between each node randomly,
and for the combiner C, we calculate it following the Metropolis
rule. Fig. \ref{fig4} shows the network structure. After 1000
iterations, the test result are showed in Fig. \ref{fig5}. We can
see that, the proposed DDMCG and DDCCG still have a better
performance than DDLMS algorithm and DDMCG is closer to the DDRLS's
curve. For the diffusion strategy, the network structure has a
significant influence on the performance of our proposed DDCG
Algorithms.

\section{Conclusions}

We have developed distributed CG algorithms for incremental and
diffusion type estimation over sensor networks. The CG- based
strategies avoid the matrix inversion and numerical instability of
RLS algorithms and have a faster convergence than LMS and AP
algorithms. Simulation results have shown that the proposed IDMCG
and DDMCG algorithms have a better performance than the LMS and AP
algorithm, and a close performance to the RLS algorithm.


\begin{thebibliography}{100}
\vspace{-0.5em}
\nonumber {\footnotesize

\bibitem{Lopes}
C. G. Lopes and A. H. Sayed, "Incremental adaptive strategies over
distributed networks", \textit{IEEE Trans. Sig. Proc.}, vol. 55, no.
8, pp. 4064-4077, August 2007. \vspace{-0.45em}
\bibitem{Lopes2}
C. G. Lopes and A. H. Sayed, ``Diffusion least-mean squares over
adaptive networks: Formulation and performance analysis",
\textit{IEEE Trans. Sig. Proc.}, vol. 56, no. 7, pp. 3122-3136, July
2008. \vspace{-0.45em}
\bibitem{Li}
L.L. Li, J.A. Chambers, C.G. Lopes,and A.H. Sayed,  ``Distributed
Estimation Over an Adaptive Incremental Network Based on the Affine
Projection Algorithm", \textit{IEEE Trans. Sig. Proc.}, vol. 58,
issue. 1, pp. 151-164, Jan. 2010. \vspace{-0.45em}
\bibitem{Cattivelli}
F. Cattivelli, C. G. Lopes, and A. H. Sayed, ''Diffusion recursive
least-squares for distributed estimation over adaptive networks,''
IEEE Trans. Sig. Proc., vol. 56, no. 5, pp. 1865-1877, May 2008.
\vspace{-0.45em}
\bibitem{Mateos}
G. Mateos, I. D. Schizas, and G. B. Giannakis, "Distributed
Recursive Least-Squares for Consensus-Based In-Network Adaptive
Estimation," IEEE Trans. Sig. Proc., vol. 57, no. 11, pp. 4583-4588,
November 2009. \vspace{-0.45em}
\bibitem{TDS_2}
P. Clarke and R. C. de Lamare, "Transmit Diversity and Relay
Selection Algorithms for Multirelay Cooperative MIMO Systems"
\emph{IEEE Transactions on Vehicular Technology}, vol.61, no. 3, pp.
1084-1098, October 2012. \vspace{-0.45em}
\bibitem{Axelsson}
O. Axelsson, \textit{Iterative Solution Methods}, New York: Cambridge Univ. Press, 1994. \vspace{-0.45em}
\bibitem{Golub}
G. H. Golub and C. F. Van Loan, \textit{Matrix Computations}, 2nd
Ed.. Baltimore, MD: Johns Hopkins Univ. Press, 1989. \vspace{-0.45em}
\bibitem{Chang}
P. S. Chang and A. N. Willson, Jr,  ``Analysis of Conjugate Gradient
Algorithms for Adaptive Filtering", \textit{IEEE Transactions on
Signal Processing}, vol. 48, no. 2, pp. 409-418, Febrary 2000.
\vspace{-0.45em}
\bibitem{delamare_jidf}
R. C. de Lamare and R. Sampaio-Neto, ``Adaptive Reduced-Rank
Processing Based on Joint and Iterative Interpolation, Decimation
and Filtering", \textit{IEEE Transactions on Signal Processing},
vol. 57, no. 7, July 2009, pp. 2503 - 2514.\vspace{-0.45em}
\bibitem{RuiSTAP2010} R.
Fa, R. C. de Lamare and L. Wang, ``Reduced-rank STAP schemes for
airborne radar based on switched joint interpolation, decimation and
filtering algorithm", \textit{IEEE Trans. Sig. Proc.}, 2010, vol.
58, no. 8, pp.4182-4194.\vspace{-0.45em}
\bibitem{Wang}
L. Wang, and R.C.de Lamare ,  ``Constrained adaptive filtering
algorithms based on conjugate gradient techniques for beamforming ",
\textit{IET Signal Processing}, vol. 4, issue. 6, pp. 686-697, Feb.
2010. \vspace{-0.45em}




}


\end{thebibliography}
\end{document}